# Detailed chemistry modelling of rotating detonations with dilute *n*-heptane sprays and preheated air

Shan Jin[a,b], Chao Xu[c], Hongtao Zheng[a], Huangwei Zhang[b,*]

[a] *College of Power and Energy Engineering, Harbin Engineering University, Harbin 150001, China*
[b] *Department of Mechanical Engineering, National University of Singapore, Singapore 117576, Republic of Singapore*
[c] *Energy Systems Division, Argonne National Laboratory, IL, 60439, USA*

______________________________________________________________________

**Abstract**

Utilization of liquid fuels is crucial to enabling commercialization of rotating detonation engines (RDE) in the near future. In this study, Eulerian-Lagrangian simulations are conducted for rotating detonative combustion with dilute *n*-heptane sprays and preheated air. Two-dimensional flattened configuration is used and a skeletal chemical mechanism with 44 species and 112 elementary reactions for *n*-heptane combustion is adopted. The flow structure, droplet distribution, and thermochemical parameters in the refill zone are first analyzed. It is shown that the mixture in the refill zone is heterogeneous, including evaporating droplets, vapor, and air. When the total temperature is below 950 K, the average equivalence ratio increases with the total temperature. When it is higher than 950 K, the average equivalence ratio is almost constant. Subsequently, the chemical explosive mode analysis is applied to identify the controlling reactions and dominant combustion modes in the fuel refill zone and reaction fronts. Results demonstrate that the initiation reaction (R104: $n$-$C_7H_{16}$ + $O_2$ → 2-$C_7H_{15}$ + $HO_2$) and low-temperature reaction (R107: $RO_2$ → R'$O_2$H) are dominant in the upstream and downstream of the refill zone, respectively. The intermediate species from low-temperature chemistry, R'$O_2$H, is found to be important for the chemical explosive mode in the undetonated mixture. The influence of species diffusion and dispersed droplets is further analyzed. Results show that vapor autoignition facilitated by droplet evaporation occurs in the refill zone. Finally, the effects of the air total temperature on the detonation propagation speed and RDE propulsion performance are investigated. It is found that the detonation propagation speed and specific impulse increase with air total temperature. The total pressure ratio first increases and then decreases as the air total temperature increases. Moreover, when the total temperature of the preheated air is above 1,300 K, the effects of low temperature chemistry are negligible.

______________________________________________________________________

*Corresponding author. E-mail:
huangwei.zhang@nus.edu.sg. Tel: +65 6516 2557.

## 1. Introduction

Rotating detonation engine (RDE) is deemed a promising pressure-gain combustion technology due to high thermodynamic cycle efficiency [1]. Fuel flexibility is crucial to materializing RDE towards a practical propulsion system. Typically, liquid fuels have high energy density and are convenient to be stored and transported. In recent years, the interests in liquid fuel RDE have revived. For instance, Bykovskii et al. successfully achieved two-phase rotating detonation waves (RDW) using kerosene sprays with oxygen-enriched air ($O_2/N_2$ = 1:1 by vol.) [2] or small addition of hydrogen or syngas [3]. They also found that hydrogen addition enables a more compact RDE combustor [3]. Kindracki [4] also performed liquid kerosene with hydrogen addition in RDE experiments. Rotating detonations were successfully achieved, with a velocity deficit of 20% − 25%, relative to Chapman-Jouguet (C-J) value. More recently, Wolański et al. [5] partially mixed preheated liquid Jet-A or gasoline with hot air, and the resultant reactant composition is higher than the rich flammability limit. They realized a rotating detonation without hydrogen addition. The RDW speed deficit is up to 35%, due to possible losses arising from chamber wall heat transfer, pre-injection deflagration, or reactant nonuniformity [5].

To shed further light on RDE with liquid fuels, numerical studies have also been available. For instance, Hayashi et al. [6] found that a steady JP-10/air RDW can be achieved within wide windows of equivalence ratio and pre-vaporization degree. They also reported that for certain JP-10 droplet concentration (e.g., 0.08-0.163 kg/m$^3$ for 3 µm droplets), detonation failure occurs. Moreover, Ren and Zheng [7] observed that under ramjet-like conditions rotating detonations with kerosene sprays (no pre-vaporization) can be achieved in a limited range of total pressure (5-7 atm) and increased total temperature is conducive to RDW stability. Besides, a bifurcated wave structure is observed near the spray injector [7].

Meng et al. [8] focused on a more volatile liquid fuel, $n$-heptane, and considered partially pre-vaporized $n$-$C_7H_{16}$ and air mixture to systematically evaluate the influences of droplet diameter (5-50 µm) and pre-vaporization degree on detonation speed and evaporating fuel droplet distribution. They found that the detonation speed decreases with decreased pre-vaporization degree or increased droplet diameter. They also analyzed the detailed gas-liquid two-phase flow structure and found that a layer with high vapor concentration exists between the droplet-laden mixture/combustion product contact surface [9]. Besides, considering the same fuel, Zhao and Zhang [10] found that the detonation propagation speed increases as the total equivalence ratio increases for the same droplet diameter. When the droplet diameter is less than 5 µm, thrust force from kinetic energy and pressure gain decreases with droplet size. Beyond 5 µm, the former first increases and then decreases with the droplet diameter, while the latter has limited change [10].

The key feature of liquid fuel RDE is that the combustion proceeds in vapor-droplet two-phase mixtures. Since preheated air is widely adopted in practical tests [5], this renders the static temperature and pressure of the undetonated mixture sufficiently high, probably inducing unexpected pre-RDW autoignition. It is shown that low-temperature chemistry (LTC) plays a significant role in autoignition process of liquid sprays in a hot atmosphere, featured by pronounced negative temperature coefficient (NTC) or zero temperature coefficient (ZTC) phenomenon [11]. However, due to the limitations of measurements and modelling approaches (simple chemistry, e.g., in [7][10]), chemical structure and reaction progress in the fuel refill zone of a liquid fuel RDE with hot air have not been studied yet. In this work, we will conduct Eulerian–Lagrangian modelling of rotating detonations with dilute liquid $n$-heptane sprays. Different from our previous work [8–10], a detailed mechanism with 44 species and 112 elementary reactions [12] will be employed for $n$-heptane combustion. The objectives of our study are to clarify: (1) the thermochemical conditions in the heterogeneous fuel refill zone of a liquid fuel RDE; (2) the chemical structures in $n$-heptane rotating detonations; (3) the effects of preheated air total temperature and LTC on detonation speed and propulsion indices.

## 2. Physical model and numerical method

### 2.1 Physical model for spray RDE

Figure 1 shows the two-dimensional computational domain of a flattened model RDE chamber. The flow structure is composed of a RDW, deflagration surface, and oblique shock. The length ($x$-direction) and width ($y$) of the domain are 81 mm and 50 mm, respectively. As annotated in Fig. 1, the outlet is non-reflective due to the local supersonic flows. Periodic conditions are enforced at the left and right sides, such that the RDW can propagate across the domain with continuous cycles.

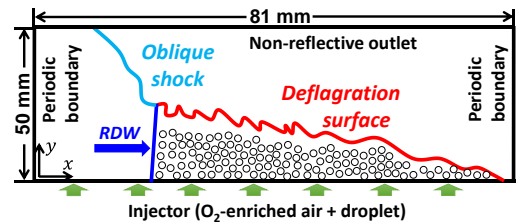

Fig. 1. Two-dimensional domain of a spray RDE.

The computational domain in Fig. 1 is discretized using 60 × 60 µm$^2$ Cartesian cells for the region of [0,

25 mm] × [0, 81 mm], whereas a cell size of 120 μm is used for the rest domain. The refined area can completely cover the RDWs in all simulations. This results in 648,000 cells for the domain in Fig. 1. Grid independence test with halved mesh size (30 μm, see supplementary document) shows that the flow and reaction structures are almost not changed.

Two-phase heterogeneous mixture is injected from a continuous inlet of the domain (at $y = 0$), including carrier gas and $n$-$C_7H_{16}$ droplets. The carrier gas is oxygen-enriched air ($Y_{O2}$:$Y_{N2} = 0.432$:$0.568$ by mass), which is deemed conducive for RDW stabilization by Bykovskii et al. [2]. Different from our previous studies [8–10], liquid fuel pre-vaporization prior to injection is not considered here and therefore $n$-$C_7H_{16}$ vapor mass fraction in the carrier gas is zero. The total pressure of the carrier gas is $p_0 = 20$ atm in all simulations. We run a series of spray RDE cases and find that when the total temperature, $T_0$, is less than 700 K, the RDW cannot stabilize. A slightly higher critical temperature (1,000 K) is reported in Ref. [7], which may be because less volatile fuel (i.e., kerosene) is considered in their simulations. In the following part of this paper, $T_0$ from 700 to 1,400 K will be studied. The gas injection follows the isentropic relations, considering the carrier gas total pressure ($p_0 = 20$ atm) and predicted gas pressure inside the chamber, and the droplet injection is synchronized with the carrier gas for both injection timing and velocity [9,10].

The droplet initial temperature is assumed to be 323 K, to mimic pre-heating by the hot carrier gas in the upstream plenum, as implemented in the RDE tests by Kindracki [4] and Wolański et al. [5]. This enables efficient gasification of fuel droplets inside the combustion chamber and therefore promote detonative combustion efficiency. The initial material density and heat capacity of liquid $n$-heptane are 680 kg/m$^3$ and 2,952 J/kg/K, respectively. Well sprayed (hence fine-grained) droplets are favorable for rotating detonations [4]. In this study, mono-sized droplets are considered, with the initial diameter $d_0 = 5$ μm. Droplet aerodynamic breakup is modelled following Ref. [13]. The liquid fuel equivalence ratio $\phi$ is used to parameterize the droplet loading, defined as the mass ratio of the droplets to the oxidant. In our simulations, $\phi = 1$ is used for all cases.

## 2.2 Numerical method

The Eulerian–Lagrangian method is used to simulate rotating detonations with sprayed liquid fuel droplets. The gas phase is described with the Eulerian method, whilst the individual fuel droplets are tracked with a Lagrangian fashion. Two-way coupling between the gas and liquid phases are implemented, considering the exchanges of mass, momentum, and energy between them. For the gas phase, the Navier-Stokes equations of mass, momentum, total non-chemical energy, and species mass fraction are solved.

For the liquid phase, the droplets are spherical and point-force approximation is adopted. The temperature gradient inside the droplets is neglected, considering their small Biot numbers (~0.057 when the droplet temperature and diameter are 323 K and 5 μm). Droplet interactions (e.g., collisions) are not considered, which is not important for dilute sprays with initial droplet volume fraction less than 0.1%. Besides, since micro-sized droplets are considered, the Basset force, lift force, and body force are not included. With these considerations, the evolutions of mass, momentum, and energy for a single droplet follow

$$\frac{dm_d}{dt} = -\dot{m}_d, \quad (1)$$
$$\frac{d\mathbf{u}_d}{dt} = \frac{\mathbf{F}_d + \mathbf{F}_p}{m_d}, \quad (2)$$
$$c_{p,d} \frac{dT_d}{dt} = \frac{\dot{Q}_c + \dot{Q}_{lat}}{m_d}, \quad (3)$$

where $t$ is time and $m_d = \pi \rho_d d^3/6$ is the mass of a single droplet, with $\rho_d$ and $d$ being the droplet material density and diameter, respectively. $\mathbf{u}_d$ is the droplet velocity vector, $c_{p,d}$ is the droplet heat capacity, and $T_d$ is the droplet temperature.

The droplet evaporation rate $\dot{m}_d$ in Eq. (1) is modelled as $\dot{m}_d = \pi d \rho_f D_f \widetilde{Sh} \ln(1 + B_M)$ [14], where $\rho_f$ and $D_f$ respectively are the density and mass diffusivity at the film over the droplet surface. The Spalding mass transfer number is $B_M \equiv (Y_{Fs} - Y_{F\infty})/(1 - Y_{Fs})$. $Y_{Fs}$ and $Y_{F\infty}$ are the fuel vapor mass fractions at the droplet surface and in the gas phase, respectively. The modified Sherwood number $\widetilde{Sh}$ is $\widetilde{Sh} = 2 + [(1 + Re_d Sc)^{1/3} \max(1, Re_d)^{0.077} - 1]/F(B_M)$, with the Schmidt number $Sc = 1.0$. The function $F(\vartheta) = (1 + \vartheta)^{0.7} \ln(1 + \vartheta)/\vartheta$ considers the variation of the film thickness due to Stefan flow effects [14]. In Eq. (2), the Stokes drag $\mathbf{F}_d$ is modelled as $\mathbf{F}_d = (18\mu/\rho_d d^2) \cdot (C_d Re_d/24) \cdot m_d (\mathbf{u} - \mathbf{u}_d)$ [15], where $\mu$ and $\mathbf{u}$ are the gas dynamic viscosity and velocity vector, respectively. The drag coefficient, $C_d$, is estimated using Schiller and Naumann model [15]. $Re_d \equiv \rho d |\mathbf{u}_d - \mathbf{u}|/\mu$ is the droplet Reynolds number, and $\rho$ is the gas density. Besides, $\mathbf{F}_p = -V_d \nabla p$ is the pressure gradient force and $V_d$ is the droplet volume.

In Eq. (3), the convective heat transfer rate $\dot{Q}_c$ is $\dot{Q}_c = h_c A_d (T - T_d)$. Here $T$ is gas temperature, and $A_d$ is the surface area of a single droplet. $h_c$ is the convective heat transfer coefficient, following Ranz and Marshall [16]. Furthermore, $\dot{Q}_{lat}$ accounts for the heat exchange rate associated with the latent heat of evaporation of liquid $n$-heptane.

The equations for the gas and liquid phases are solved using a customized OpenFOAM code, *RYrhoCentralFoam*. The solver is carefully validated and verified for shock capturing, molecular diffusion, flame-chemistry interactions and gas-liquid two-phase problems [17–19]. For gas phase, second-order backward scheme is used for time marching, and the time step is about $2 \times 10^{-9}$ s. Second-order Godunov-type upwind-central scheme is employed to calculate the convection fluxes in the momentum equations. The total variation diminishing scheme is applied for the convection terms in energy and species equations. A detailed mechanism (44 species and 112 reactions) [12] is used. Low-temperature chemistry is included in this mechanism, and hence the LTC effects and two-stage ignition in *n*-heptane rotating detonations can be studied. One simulation with 88 species and 387 reactions (see details in supplementary document) is also run and the results demonstrate that differences between these two mechanisms are negligible.

For the liquid phase, fuel droplets are tracked from the barycentric coordinates. Equations (1) − (3) are integrated with first-order Euler method and the right-hand-side terms are treated in a semi-implicit approach. Details of the numerical method in *RYrhoCentralFoam* are available in [17–19].

The simulations are run on the ASPIRE 1 Cluster from National Supercomputing Centre in Singapore and 360 processors are used for each case. The simulated physical time is about $0.62 - 0.71$ ms, corresponding to 15 cycles.

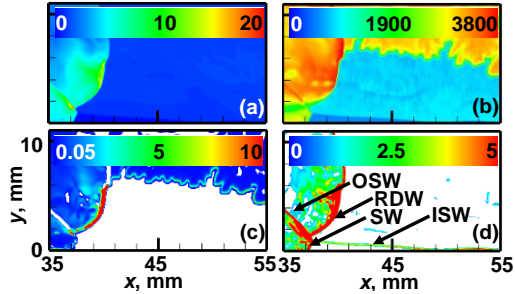

Fig. 2. (a) Pressure (in MPa), (b) gas temperature (K), (c) heat release rate ($\times 10^{12}$ J/m$^3$/s), and (d) pressure gradient magnitude ($\times 10^9$ Pa/m). $T_0 = 1,000$ K.

## 3 Results and discussion

### 3.1 Thermochemical condition

Figure 2 shows the distributions of pressure, gas temperature, Heat Release Rate (HRR), and pressure gradient magnitude after the single-waved rotating detonation stabilizes (over ten cycles). The carrier gas total temperature is $T_0 = 1,000$ K. The basic structure of a RDE flow field is well captured. As seen from Fig. 2(b), the temperature in the fuel refill zone (enclosed by the RDW, deflagration surface, and inlet) varies between 510 and 950 K, and the mean is about 822 K. Besides, the pressure in the refill area is 1.78 atm – 11.5 atm with a mean of 10.68 atm. A notable feature in Fig. 2(c) is that although HRR is high along the RDW, detonative combustion does not maintain near the injector and only a shock wave can be found (marked as SW in Fig. 2d). A multi-wave structure can also be seen, connecting the induced shock wave (ISW), oblique shock wave (OSW) and the SW, as shown in Fig. 2(d). The ISW is induced by the high-speed injection of the air, which is also found in Ref. [20]. The gas between the ISW and the injector is featured by high speed, low temperature, and pressure. This therefore leads to slower propagation of the SW compared to the RDW. This also affects the heating, evaporation and movement of the liquid droplets in this area, which will be discussed in Fig. 3.

Plotted in Fig. 3 are the distributions of dispersed fuel droplets in the refill zone, and they are colored by their *y*-component velocity, temperature, evaporation rate, and diameter. After being injected into the combustor, the velocities of *n*-heptane droplets near the injector are relatively high before the ISW, up to 1,000 m/s, and they gradually relax towards about 375 m/s. Meanwhile, as seen from Fig. 3(b), the droplets are quickly heated to their saturation temperature and start to vaporize. The droplet evaporation rate from Fig. 3(c) gradually increases (due to increased droplet temperature), peaks at 2 mm, and then decreases (due to decreased droplet mass) along the *y*-direction inside the refill zone.

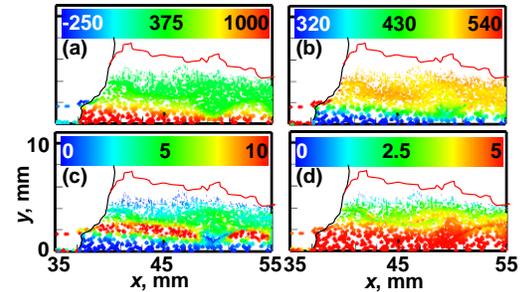

Fig. 3. Droplet (a) *y*-component velocity (in m/s), (b) temperature (K), (c) evaporation rate ($\times 10^{-9}$ kg /s), and (d) diameter (μm). $T_0 = 1,000$ K. Black line: detonation and shock waves; red line: deflagration surface.

Figures 4(a) and 4(b) respectively show the distributions of the effective equivalence ratio, $\phi_{eff}$, when the detonation wave propagates steadily at $T_0 = 700$ K and 1,000 K. In this work, $\phi_{eff}$ is calculated from the ratio of required stoichiometric oxygen atoms to the available oxygen atoms [21], i.e., $\phi_{eff} = 2(n_C + n_H/4)/n_O$, where $n_C$, $n_H$, and $n_O$ denote the number of available carbon, hydrogen and oxygen atoms, respectively. Be reminded that since it is based on element conservation, $\phi_{eff}$ is also well defined in the product gas. However, the ones in the un-detonated mixtures (e.g., fuel refill zone) are relevant for our analysis. One can see from Fig. 4 that, very limited vapor (blue areas) exists near the injectors into the RDE chamber, and the starvation of fuel vapour leads to local detonation decoupling near the injector, as shown in Fig. 2(d).

As the air total temperature increases, more fuel vapor is present near the injector, through comparing Figs. 2(a) and 2(b). The variation of averaged effective equivalence ratio $\phi_{eff}$ under different $T_0$ is given in Fig. 4(c). The average static temperature $T_{re}$ in the fuel refill zone at different $T_0$ is also given below the x-axis in Fig. 4(c). Here we average $\phi_{eff}$ respectively based on: (1) the refill zone (based on the criterion of local temperature lower than the corresponding $T_0$) and (2) detonation wave front (HRR > $10^{13}$ J/m$^3$/s [22]). As $T_0$ increases from 700 K to 950 K, the average static temperature of the gas in the refill zone increases from 581 K to 787 K, the average $\phi_{eff}$ increases from 0.668 to 0.69. However, when $T_0$ further rises from 950 K to 1,300 K, the average $\phi_{eff}$ in the refill zone is almost constant, about 0.69, which is slightly higher than the lean flammability limit of n-heptane, 0.56 [23].

For the average $\phi_{eff}$ at the RDW, when $T_0$ < 950 K, a portion of liquid droplets cannot fully vaporize ahead of RDW, resulting in a relatively low $\phi_{eff}$ at the RDW in Fig. 4(c). As $T_0$ further increases, the remaining droplets after the RDW gradually decreases, and hence the equivalence ratio at the RDW is close to unity, which is the ER of the injected mixture in our simulations. Therefore, $T_0$ = 950 K is a critical total temperature for the ERs for the fuel vapor availability at the detonation wave. If we linearly extrapolate the three points of $\phi_{eff}$ (RDW) to 600 K, the corresponding $\phi_{eff}$ at the RDW is about 0.62, which is near the lean flammability limit. This also justifies why a stable RDW cannot be achieved with $T_0$ < 700 K under the simulated conditions.

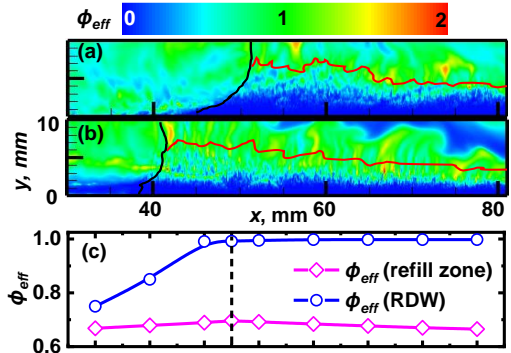

Fig. 4. Contours of effective equivalence ratio with $T_0$ = (a) 700 K and (b) 1,000 K. (c) Average effective equivalence ratio inside the fuel refill zone and at the detonation wave as functions of total temperature. Black line: detonation/shock waves; red line: deflagration surface. $T_{re}$: average static temperature in the refill zone.

### 3.2 Chemical structure

Here the chemical reaction characteristics in the spray RDE will be extracted with the chemical explosive mode analysis (CEMA) [24]. The equation for a gas reaction system reads:

$$\frac{D\boldsymbol{\omega}(y)}{Dt} = \mathbf{J}_{\omega} \frac{D\boldsymbol{y}}{Dt} = \mathbf{J}_{\omega}(\boldsymbol{\omega} + \boldsymbol{s} + \boldsymbol{d}), \quad \mathbf{J}_{\omega} = \frac{\partial \boldsymbol{\omega}}{\partial y} \quad (4)$$

where $D(\cdot)/Dt$ is the material derivative, $\boldsymbol{y}$ is the vector of temperature and species concentrations, $\boldsymbol{\omega}$ is the chemical reaction, $\boldsymbol{s}$ is the diffusion term, $\boldsymbol{d}$ is the droplet evaporation term and $\mathbf{J}_{\omega}$ is the Jacobian of the chemical system. A chemical explosive mode (CEM) is defined as the eigen-mode associated with a positive (real part) eigenvalue which indicates that the local mixture tends to ignite under lossless conditions [24]. Distributions of the CEM eigenvalue $\lambda_e$ are shown in Fig. 5 for $T_0$ = 1,000 K. For better illustration, the logarithmic expression of the eigenvalue $\lambda_e$ is plotted, i.e., $\lambda_{CEM} \equiv \text{sign}[\text{Re}(\lambda_e)] \cdot \log_{10}[1 + |\text{Re}(\lambda_e)|]$. Details of the CEMA can be found in [24].

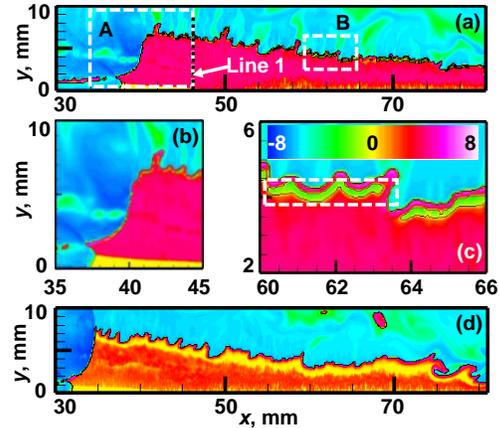

Fig. 5. Distributions of (a) CEM eigenvalue, close-up view of (b) zone A and (c) zone B, and (d) CEM eigenvalue without low-temperature chemistry for $T_0$ = 1,000 K.

Evident from Fig. 5(a) is that the heterogeneous two-phase mixture in the refill zone features large positive eigenvalues, indicating that fast droplet evaporation and vapor/air mixing turn the gas into chemically explosive state. Zero-crossing of the eigenvalue occurs at the detonation and deflagration surfaces. As shown in Fig. 5(b), ahead of the RDW, as the droplets are just sprayed into the chamber, the evaporation rate is low and the n-$C_7H_{16}$ vapor cannot mix effectively with the air, which results in overall fuel-lean composition (local ERs: 0–0.03) and the eigenvalue here is around zero. Near the deflagration surface shown in Fig. 5(c), some striped burned zones appear, with small negative eigenvalues (marked as dashed box). The reason for it will be further discussed in Fig. 6. For comparison, one additional simulation is run, in which the low-temperature elementary reactions, i.e., R105-R112 (Appendix A of [12]), are deactivated. Distribution of the CEM eigenvalue $\lambda_{CEM}$ from this test is shown in Fig. 5(d). In the refill zone, $\lambda_{CEM}$ is around 2, much lower than that of Fig. 5(a), i.e., about 4. This implies that the chemical timescale predicted with LTC included is approximately two orders of magnitude shorter than that without LTC. This clearly shows the LTC promotes the overall reactivity of the undetonated gaseous mixtures.

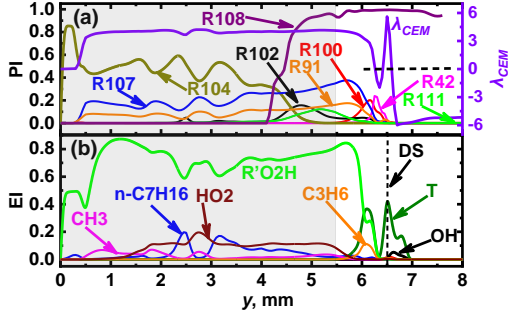

Fig. 6. Distributions of (a) PI and CEM logarithmic eigenvalue, and (b) EI at line #1 in Fig. 5. Shaded zone: droplet-laden area; DS: deflagration surface.

The chemical reactions in liquid $n$-heptane RDE will be further analyzed through the profiles of the participation index (PI) and explosion index (EI) [24] across the refill zone and the deflagration surface at $x$ = 45 mm (annotated as line 1 in Fig. 5a). High PI (EI) signifies the dominance of the corresponding elementary reaction (species) in the explosive mode. The results are shown in Fig. 6. The curve of $\lambda_{CEM}$ along line 1 is also plotted in Fig. 6(a). As seen from Fig. 6(a), the gas reactivity is weak ($\lambda_{CEM}$ being very low, but still positive) near the injector, at $y < 0.5$ mm. This is because very limited droplet evaporation and the low gas temperature (about 520 K) in the previously mentioned small region before the ISW. Beyond $y = 0.5$ mm, $\lambda_{CEM}$ rises quickly and then levels around 4.0. Besides, at $y = 0$-5.5 mm (i.e., the shaded area, with heterogeneous mixtures of fuel droplets, vapor, and air), the initiation reaction R104 ($n$-$C_7H_{16} + O_2 \rightarrow 2$-$C_7H_{15} + HO_2$) is significant with the highest PI. After that, R107 ($RO_2 \rightarrow R'O_2H$), R111 ($OR''O_2H \rightarrow OR''O + OH$) and R91 (1-$C_7H_{15}$ $\rightarrow$ 1-$C_5H_{11} + C_2H_4$) become dominant. The first two reactions are low-temperature reactions and last one is a cracking reaction. For these locations, high EI of R'$O_2$H is also observed, which is an intermediate species generated and consumed by the low-temperature reactions, R107 and R108 (R'$O_2$H + $O_2$ $\rightarrow$ $O_2R'O_2H$), respectively.

Beyond the droplet-laden area (i.e., $y > 5.5$ mm), since the droplet evaporation is completed, the local mixture is gaseous (air and $n$-heptane vapor). The PIs for the LTC (e.g., R107) decrease. Instead, the PIs of the following elementary reactions become comparatively high: R42 ($CH_3 + HO_2 \rightarrow CH_3O + OH$), R100 ($n$-$C_7H_{16} + OH \rightarrow 2$-$C_7H_{15} + H_2O$) and R108 become dominant. At $y = 6.35$ mm, first zero-crossing of the $\lambda_{CEM}$ curve can be found, corresponding to a high temperature EI (indicative of thermal runaway). Burned mixture next to it features $\lambda_{CEM} < 0$. This corresponds to the first-stage ignition

(i.e., green areas in Fig. 5c), which produces small radicals such as $C_3H_6$ or $C_2H_2$. Further downstream, the mixture becomes explosive again before it gets burned in the second-stage ignition near the reactant-product contact surface, with the $\lambda_{CEM}$ being much higher than that before the first-stage ignition. This can also be corroborated from the EI of temperature in Fig. 6(b), and is consistent with the findings for $n$-heptane autoignition in Ref. [25]. Nonetheless, differently, our results indicate that thermal runaway are significant for both stages, which is because of sufficient radical runaway in both two-phase and gas-only mixtures in the refill zone.

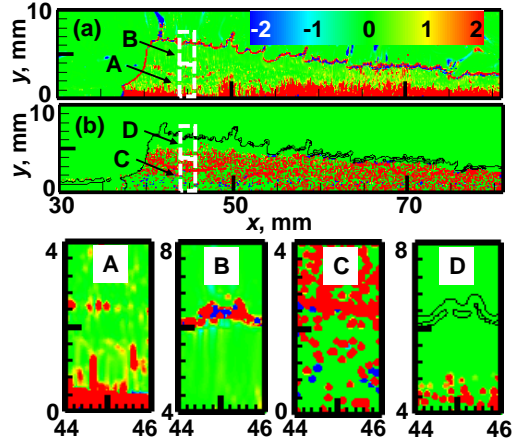

Fig. 7. Distributions (a) $\alpha_s$ and (b) $\alpha_d$ and their enlarged views. $T_0$=1,000K. Black line in (b): isolines of $\lambda_{CEM} = 0$. A−D correspond to the boxes in Figs. (a) and (b).

The $n$-heptane vapor ignition modes in the refill zone are further analyzed in Fig. 7. Here the ignition mode is identified by projecting $D\boldsymbol{\omega}(y)/Dt$ to the CEM using the left eigenvector $\boldsymbol{b}_e$ [26]:

$$\frac{D\phi_\omega}{Dt} = \lambda_e\phi_\omega + \lambda_e\phi_s + \lambda_e\phi_d + \frac{D\boldsymbol{b}_e}{Dt}\cdot\boldsymbol{\omega}, \quad (5)$$

where $\phi_\omega = \boldsymbol{b}_e \cdot \boldsymbol{\omega}$, $\phi_s = \boldsymbol{b}_e \cdot \boldsymbol{s}$ and $\phi_d = \boldsymbol{b}_e \cdot \boldsymbol{e}$ respectively represent the projected chemical, diffusion and evaporation terms. Note that here $\phi_d$ include both the thermal affect for the temperature and the kinetic effect for the vapor. The last term in Eq. (5) can be neglected, following Refs. [26]. The effects of species diffusion and droplet evaporation on gas reactions can be indicated respectively by the ratios of $\alpha_s = \phi_s/\phi_\omega$ and $\alpha_d = \phi_d/\phi_\omega$. If $\alpha > 1$, ignition is facilitated by diffusion or droplet evaporation; if $|\alpha| < 1$, chemistry is dominant (hence autoignition); if $\alpha < -1$, diffusion or evaporation dominates chemistry and inhibits ignition [26].

The distributions of $\alpha_s$ and $\alpha_d$ in the refill zone are shown in Fig. 7. It can be seen from Fig. 7(a) that $\alpha_s$ is high (in red) near the spray injector in the entire refill zone, which can be clearly seen from the close-up view of zone A in Fig. 7(A). This indicates that the ignition of the local vapor is promoted by the species diffusion. This is understandable since in this region the equivalence ratio is still very low (see Fig. 4b), and efficient species diffusion (hence reactant mixing) is favorable for vapor ignition. Further downstream, e.g., $y > 5$ mm, the ignition mode changes to auto-ignition ($\alpha_s \approx 0$) and fuel diffusion plays a limited role. In these areas, the reactant composition is overall uniform, although reactant stratification still exists due to discrete distributions of the evaporating droplets. Near the deflagration surface (zone B), although diffusion is dominant, mixed local combustion modes are present in Fig. 7(B); the reactions at some pockets are inhibited by the species diffusion.

Likewise, how droplet evaporation affects vapor reaction ahead of the rotating detonation wave can be examined through the distributions of $\alpha_d$. At positions near the injector, the droplets are just sprayed into the chamber and cannot evaporate quickly, which results in small $\alpha_d$, as shown in Figs. 7(b) and 7(C). As the droplet evaporation accelerates, its contribution towards the CEM becomes high (red spots in Fig. 7C) and is important for the entire droplet-laden refill zone. Nonetheless, some sparse blue dots are observed, which is probably due to the heat absorption by the evaporating droplets. Moving further downstream towards the deflagration front (Fig. 7D), the evaporation affect becomes negligible as the droplets are fully evaporated.

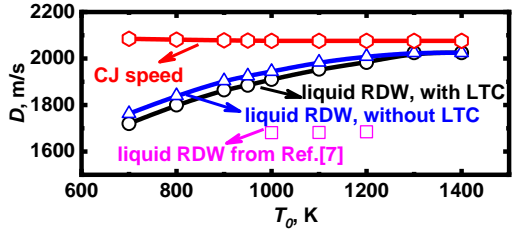

Fig. 8. Detonation wave speed versus total temperature.

### 3.3 Carrier gas total temperature effects

Figure 8 shows the change of average detonation wave speed $D$ with different air total temperature $T_0$. Here $D$ is calculated from the pressure history at a probe near the head end of the domain. The Chapman–Jouguet speeds of gaseous n-heptane/air mixtures with $\phi = 1.0$ are added. The liquid fuel RDW speeds are 3%−15% lower than the C-J values. This may be caused by, e.g., imperfect reactant mixing and droplet evaporation [5][10]. It is also found that the RDW speed with liquid fuel increases when $T_0$ is increased, which is also observed by Meng et al. [9]. This is because the increase of $T_0$ raises the temperature in the fuel refill zone and increases the rate of mixing between reactants which and ultimately causes an increase in $D$. This trend is different from the result of Ref. [7], because the droplets in [7] are finer (2 μm), which enables complete evaporation before the RDW arrives and therefore the equivalence ratio of the undetonated gas has weak dependence on $T_0$. In addition, the average speeds predicted without LTC are higher than those with LTC. This is because the low-temperature chemical reactions in the fuel refill zone consumes part of the gaseous n-$C_7H_{16}$.

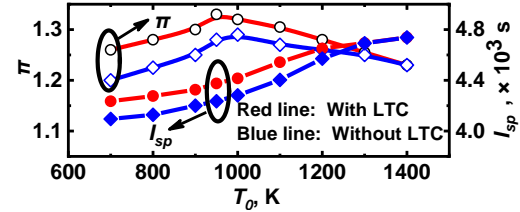

Fig. 9. Specific impulse and total pressure ratio versus total temperature.

Figure 9 shows the effects of air total temperature on the total pressure ratio $\pi$ and specific impulse $I_{sp}$. The pressure ratio is $\pi = p_3/p_0$, where $p_3$ is the total pressure from near the exit ($y = 49$ mm) and $p_0$ is air injection total pressure, i.e., 20 atm in this study. The specific impulse is calculated following Ref. [10]. One can see that the pressure ratio first increases and then decreases with increasing $T_0$. From $T_0 = 700$ K to 950 K, an increase in $T_0$ changes the effective equivalence ratio at the RDW (see Fig. 4), which causes a gradual increase in $p_3$ and ultimately an increase in $\pi$. From 1,000 K to 1,400 K, the average equivalence ratio at the RDW front no longer changes with increasing $T_0$. However, as $T_0$ increases, chemical reactions accelerate, the deflagrative combustion intensifies and thus lowers the detonated fuel fraction and $p_3$, which eventually leads to a decrease in the pressure ratio (the detonated fuel fraction versus total temperature is shown in supplementary document for interested readers).

Nonetheless, the specific impulse monotonically increases with $T_0$. This is because injection with higher air total temperature expands the area of the low-temperature region before the ISW mentioned in Section 3.1. The gas velocity in the mentioned region is high (more than 1,000 m/s). The increase of the area increases the gas velocity and eventually increases the specific impulse.

The results predicted without LTC are also shown. One can find that the $I_{sp}$ without LTC is consistently underpredicted. This is because the deactivation of the LTC changes CEM eigenvalue $\lambda_e$ (lower reactivity, see Fig. 5) in the refill zone. The change in the CEM eigenvalue results in a lower detonated fuel fraction in the detonation combustion (see supplementary document), thereby a lower pressure of the detonation wave and ultimately to a lower specific impulse and pressure ratio. When $T_0 > 1,300$K, the propulsion indices are almost not affected by the LTC, because the LTC is inhibited for this temperature range.

## 4 Conclusion

Two-dimensional rotating detonations with liquid *n*-heptane sprays and preheated air are simulated with a Eulerian-Lagrangian method and a skeletal chemical mechanism. The results show that the mixture in the refill zone is heterogeneous, including evaporating droplets, vapor, and air. When the total temperature is below 950 K, the average equivalence ratio increases with the total temperature. The chemical structures in the refill zone and reaction front are studied with the chemical explosive mode analysis. It is seen that with fuel vapor addition and efficient mixing, the mixture becomes explosive in most of the refill zone. The initiation reaction (R104) and low-temperature reaction (R107) are dominant in the upstream and downstream of the refill zone, respectively. The LTC intermediate species, R'O$_2$H, is found to be important for chemical explosive mode in the undetonated mixture. The influence of species diffusion and dispersed droplets on fuel vapour ignition is further analyzed. The detonation propagation speed and specific impulse increase with air total temperature. The total pressure ratio firstly increases and then slightly decreases. Inclusion of the LTC in the chemical mechanism would affect the predictions of these parameters, but the difference is minimized when the air total temperature is above 1,300K.